# Aipan VR: A Virtual Reality Experience for Preserving Uttarakhand's Traditional Art Form

Nishant Chaudhary[1], Mihir Raj[1], Richik Bhattacharjee[2], Anmol Srivastava[1], Rakesh Sah[1], Pankaj Badoni[2]

{nishantchaudhary163@gmail.com, mihirraj346@gmail.com, richik2000@hotmail.com, asrivastava@ddn.upes.ac.in, rakesh.sah@ddn.upes.ac.in, pbadoni@ddn.upes.ac.in}

School of Design[1],
School of Computer Science[2],
University of Petroleum and Energy Studies, Uttarakhand, India

**ABSTRACT**

This paper presents a demonstration of the developed prototype showcasing a way to preserve Intangible Cultural Heritage of Uttarakhand, India. Aipan is a traditional art form practiced in the Kumaon region in the state of Uttarakhand. It is typically used to decorate floors and walls at the places of worship or entrances of homes and is considered auspicious to begin any work or event. This art is associated with a great degree of social, cultural as well as the religious significance and is passed from generation to generation. However, in the present era of modernization and technological advancements, this art form now stands on the verge of depletion. This study presents a humble attempt to preserve this vanishing art form through the use of Virtual Reality (VR). Ethnographic studies were conducted in Almora, Nainital, and Haldwani regions of Uttarakhand to trace the origins as well as to gain a deeper understanding of this art form. A total of ten (N = 10) aipan designers were interviewed. Several interesting insights are revealed through these studies that show the potential to be incorporated as VR experience.

**Keywords**: Intangible Cultural Heritage, Virtual Reality, Aipan, Uttrakhand

## 1 INTRODUCTION

'Aipan' originates through the Sanskrit word 'Arpan' (dedication). A type of traditional art associated with fortune and fertility resembles making an offering to God. Aipan designs are depicted primarily in venerable locations, along with the house's main entrance door and front courtyard. To begin with, every auspicious work or occurrence is a sign of goodness. As described by the Ministry of Culture, India [1], "*this artform mainly comprises of geometric patterns involving lines, dash, dot, circle, square, triangle, swastika and lotus, all of which seem to have had their origin in the Puranas and Tantric rituals*". Aipan has also been declared among one of the Intangible Cultural Heritage of India under the domain of Traditional Craftsmanship and is required to be protected. As this art form is mainly practiced in rural areas of Kumaon region of Uttarakhand, it faces several challenges which can be broadly categorized under demographic issues, weakened practice and transmission and loss of objects or systems categories as defined by UNESCO [2]. This study presents an attempt to preserve this diminishing artform of Aipan and raise awareness among the community through the use of VR. Published studies on VR [3] for cultural and heritage preservation indicate a positive potential of this technology for experiencing culture and improving accessibility to areas that are physically constrained. As Aipan is predominantly practiced in rural hilly areas of Uttarakhand, VR acts as a suitable medium for experiencing these locations and learning the craft of Aipan. Another important aspect that our demonstration captures is showcasing the tacit and ancestral knowledge behind aipan making process and their significance. Thus also adding an educational value to the overall experience.

### 1.1 Anatomy of Aipan

Aipan art is a daily practice in some houses, where simple design can be done for ordinary days and elaborate designs are prepared on ritual and festive occasions.

- The art is practiced near Tulsi (Basil) pot/ platform.
- At the entrance of the House/door step. To welcome the guest with grace and elegance.
- On the walls to decorate the house.
- Inside the Pooja room, because these patterns are considered as symbols of good omen, which evoke divine power and brings good fortune for the family.

Following are the types of chauki and creepers used in Aipan and These insights were translated and encapsulated in our VR experience.

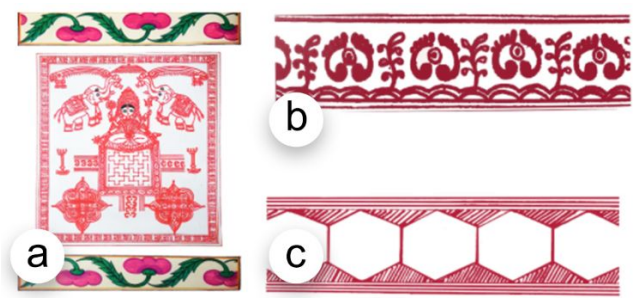

Fig 1. Different types of chauki and creepers used in Aipan. (a) Mahalaxmi chauki*. (b, c) Creepers: Laxmi paer and chota himachal.[7]

Fig 1(a) depicts a figurative drawing made for the worship of Mahalaxmi goddess, especially on a bronze plate and bamboo sticks are used in place of fingers. The colors used are red or a specific red color made from turmeric and lemon juice by a definite process. The elephant is symbolic of prosperity, water, and clouds. Laxmi is seated on the Saraswati Chauki and in the bottom are drawn two 8 cornered water pools. Fig 2 (b,c) represent various creepers. Fig 1 (b) is called Laxmi paer , i.e. footprint of goddess Laxmi. Fig 1 (c) is called chhota himachal, i.e. Valley of high range of Himalaya.

---

*Mahalaxmi Chauki image adapted from a local artist during ethnographic survey

## 2 EXPLANATION OF DEMONSTRATION

The VR demonstration comprises various Aipan art forms and its relevant elements such as chowki, creeper and the making process. The main museum building is a rough three-dimensional (3D) replica of a famous building of an Aipan artist in Almora (see Fig 2. (c)). The experience consists of users walking across various aipan artworks and getting to know their significance and use in various holy practices (refer to Fig 2.). Emphasis has been put on capturing the cultural essence of the Kumaon region through background music, aesthetic elements of museum building and the natural surroundings. To further familiarize the users with the place, a famous market has been replicated in 3D within the museum. These users can navigate across this market to further imbibe the essence of the rural hilly region that is often composed of souvenirs and products that incorporate aipan art form. Fig 3, depicts the screenshots of various elements of museum experience.

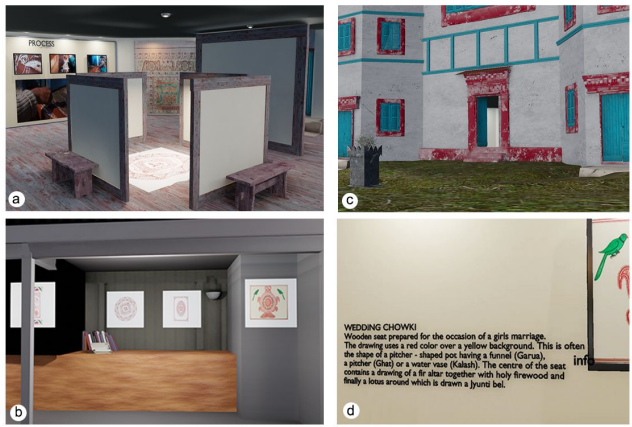

Fig 2. Screenshots of Aipan VR museum experience. (a, b) The museum displays various images and the process of making aipan. (c) Main museum building captures contextual information and environment. (d) Text and voice based information is presented to the user.

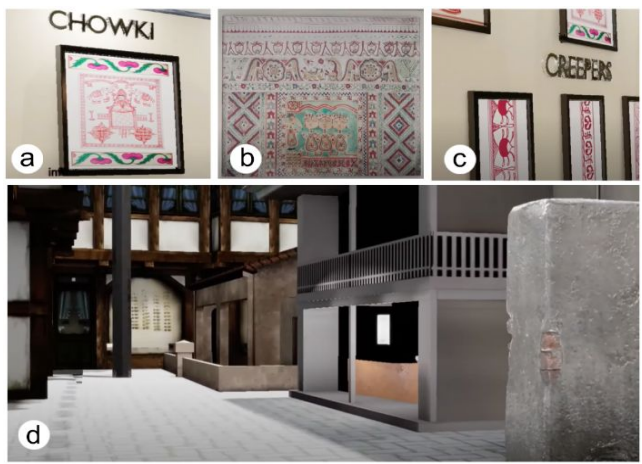

Fig 3. Elements of VR museum. (a,b,c) Various elements of aipan art form and craft have been presented. (d) Famous market places have been replicated in 3D and clubbed with the museum to help users navigate and get the essence of the location also.

## 3 WHAT MAKES IT UNIQUE AND SPECIAL

Almora is the cultural capital of Kumaon famous for its alluring beauty, Panoramic view of the Himalayas, rich cultural heritage, unique handicraft, and delicious cuisine. VR has the power to transport users to places they might never be able to visit in real life so welcoming digital visitors into the museums of the world is a natural fit. It's also a huge win for students and researchers across the globe as they will be able to explore the ritualistic folk art of Kumaon and the culture of Almora.

## 4 EXPLANATION OF THE NOVELTY

In our research we observed how this craft of Kumaon was depleting, so we tried to give it a new platform through VR. The literature survey also indicates a dearth of published articles on the Aipan artform, especially in terms of bringing in new technologies. This demonstration presents a first step towards digitizing and experiencing the Aipan craft through VR. The demonstration also attempts to utilize this experience for educational purposes. The information presented to the users is available in three audio language formats - English, Hindi, and Kumaoni.

## 5 WHY WILL IT DRAW THE CROWD

It will draw a crowd as the VR experience will help them in saving their time and money in exploring the culture. VR will also provide a unique perspective and make these experiences more accessible and realistic.

**Walkthrough video -** https://youtu.be/ZCMYe45pU5l

**Tools Used for prototyping VR**

Blender [4], Unity 3D [5], Adobe Photoshop [6]

*Mahalaxmi Chauki image adapted from a local artist during ethnographic survey